\documentclass[times]{elsarticle}

\usepackage{prletters}
\usepackage{framed,multirow}
\usepackage{subfigure}

\usepackage{natbib}
\usepackage[justification=centering]{caption}

\begin{document}

\begin{frontmatter}




\title{An Analysis of Frequent Patterns in the World Trade Web}


\author[a]{Maddalena D'Anna\corref{cor1}}
\author[a]{Alfredo Petrosino}

\address[a]{Department of Science and Technologies, University of Naples ``Parthenope'', Isola C4, Centro Direzionale, 80143 Napoli, Italy}

\begin{abstract}
This paper employs a weighted network approach to study the empirical properties of the web of trade relationships among world countries, and its evolution over time. We show that most countries are characterized by weak trade links; yet, there exists a group of countries featuring a large number of strong relationships, thus hinting to a core-periphery structure. The World Trade Web (WTW) is characterized by the following representation: a  directed graph connecting world Countries with trade relationships, with the aim of finding its topological characterization in terms of motifs and isolating the key factors underlying its evolution. 
Frequent patterns can identify channels or infrastructures to be strengthened and can help in choosing the most suitable message routing schema or network protocol. 
In general, frequent patterns have been called {\it motifs} and overrepresented motifs have been recognized to be the low-level building blocks of networks and to be useful to explain many of their properties, playing a relevant role in determining their dynamic and evolution. 
In this paper triadic motifs are found first partitioning a network by strength of connections and then analyzing the partitions separately.  The WTW has been split based on the weights of the graph to highlight structural differences between the big players in terms of volumes of trade and the rest of the world. As test case, the period 2003-2010 has been analyzed, to show the structural effect of the economical crisis in the year 2007.
\end{abstract}

\begin{keyword}
 \sep World Trade Web \sep frequent patterns \sep motif analysis




\end{keyword}
\cortext[cor1]{Corresponding author: }
\ead{maddalenadanna.uniparth@gmail.com}
\end{frontmatter}





\section{Introduction}
\label{main}

In the last decades, a large body of empirical contributions have increasingly studied socio-economic systems in the framework of network analysis.1 A network is a mathematical description of the state of a system at a given point in time in terms of nodes and links. The idea that real-world socio-economic systems can be described as networks is not new in the academic literature (Wasserman and Faust, 1994). Indeed, sociologists and psychologists have been employing social network analysis since the beginning of the last century to explore the patterns of interactions established among people or groups (Freeman, 1996; Scott,
2000). 
While any definition is necessarily imprecise, the \lq\lq Internet of Things\rq\rq\   (IoT from now on) commonly refers to objects (Things) of everyday use that have the ability to communicate among themselves and to acquire, share and process data from the surrounding environment. Considering that the IoT enabled devices are rapidly reducing their cost and size, at the same time increasing their computing power, it is easy to forecast an exponential growth in the number of   
embedded systems able to communicate and in the software applications able to exploit them (see \cite{Atzori} for a survey). 
Necessary conditions for each Thing in the IoT are unique identifiers, high level communication protocols, high level abstractions of the automation possibilities of each device and widespread standards for representing, storing and processing harvested data. 
One of the crucial aspect of IoT that affects the traffic load and the effectiveness of communication among devices  is the communication model, for which four different proposal emerged in the literature:
\begin{itemize}
\item  device to device;
\item device to cloud;
\item device to gateway;
\item back-end data sharing.
\end{itemize}
Whatever the chosen model is, the actual communication pattern can in general be represented as a directed graph, whose nodes represent the \lq\lq Things\rq\rq\  and whose directed edges represent the sent messages. 
Interestingly enough, in case of the device to device communication model, things communicate over a network that mimics the topology of social networks or human organizations.
In such a scenario, frequent patterns of interaction among nodes can identify channels or infrastructures to be strengthened and can help in choosing the most suitable message routing schema or network protocol. The effectiveness of communication ultimately depends on the network topology and on the chosen protocol.

Given a Network,  the frequent patterns of interaction among
nodes, that is overrepresented subgraphs of predefined dimensions, are called {\it motifs}: there are 13 possible motifs on directed graphs with 3 nodes and 199 possible motifs of 4 nodes. 
Small motifs (size 3 or 4) can be considered as building blocks of complex networks and are useful
indicators of the local structure of the network, often helping to explain many of the general
properties of the studied system. It has been observed~\cite{milo2002network} that not all functional motifs
playing a role in network dynamics or evolution are statistically overrepresented, but they
are nonetheless relevant to highlight local interaction patterns that characterize the
network. To evaluate if a motif is statistically significant it is common to compute a $z$-score
and the corresponding $p$-value randomizing the network in a way that preserves as much
as possible its structure, that is generating many graphs similar to the original and
evaluating the empirical distribution of each pattern. Well known problems to be addressed
to find motifs are due essentially to the broad search space to be explored, with
consequent impractical slowness, especially for higher order subgraphs: subgraph
isomorphism is a $NP$-hard problem and the number of subgraphs to be analyzed grows
exponentially with graph and motif size. To cope with the broadness of the search space
probabilistic algorithms and alternative measures of frequency, allowing or preventing
node and edge overlap, can be used.
The frequency of a motif is by definition the number of times that the motif is found as
subgraph of a given network. While counting matches, nodes and edges overlap can be
allowed or prevented, so three different measures of frequency exist: $F1$ that allows
overlap of nodes and edges arbitrarily; $F2$ that allows node overlap and prevents edge
overlap; $F3$ that prevents any overlap of nodes and edges among subgraphs.
To estimate empirically the frequency of each motif it is necessary to generate a
number of random networks, so a suitable algorithm for randomization must be provided:
the most common is edge switching among nodes, keeping constant the directionality, the
input degree and the output degree of each node. A detailed survey on all these facets can be found in~\cite{kim2013network,wong2010network}.
Application domains for the motifs search are countless, just to name a few: gene regulatory networks, PPI (protein-protein interactions) graphs and neuronal connectivity graphs in biology and
biochemistry; food networks in ecology; electronic circuits and routing algorithms in engineering; the World
Trade Web (WTW from now on) in economics. Among others, the WTW is of peculiar
interest since it is the graph obtained considering as nodes the Countries of the World
Trade Organization and as weighted edges the import/export relationships among
Countries, with billions of US dollars as the unity of weight for the total exchange volume. It is a well
studied graph, prototype of a complex network, whose topological properties have been
explored explicitly since the early 2000 years~\cite{chiarucci2014detecting,squartini2012triadic,saracco2015detecting,fagiolo2010evolution,li2003complexity,barabasi2000scale,squartini2011randomizing,ruzzenenti2012spatial,serrano2007patterns,garlaschelli2007interplay,serrano2003topology,skowron2014spanning}
and that is known to be scale-free,
with a power law distribution of the degrees of the nodes.
In the following, the search for overrepresented three nodes patterns in two
subgraphs obtained splitting the export WTW based on the strength of the connections will
be considered, limited to the period 2003-2010, trying to highlight the structural effect of
the economical crisis in the year 2007.

\section{Algorithm}
\label{A}

In order to count the exact frequency of a network motif of size $k$ it is necessary first
to enumerate all subgraphs of size $k$ in the network, then to group them into isomorphic
classes. After that, a random graph model is needed to evaluate if a motif class is
overrepresented.
The fast RAND-ESU algorithm proposed in~\cite{wernicke2005faster} is used hereafter: it implements an
efficient strategy to enumerate all size-$k$ subgraphs building a recursive tree called an
ESU-tree for each node in the network.
Assuming that all nodes are labeled with the integers $1,\dots,n$ and that the
neighbourhood of a set of nodes $\{C_{i}\}$ is the set of nodes $\{N_{j}\}$ adjacent to at least one node
in $\{C_{i}\}$, the algorithm ESU (Enumerate SUbgraps) builds the tree recursively for each node,
starting from one node and adding to it all nodes (up to size $k$) that have two properties:
the index of the added nodes must be larger than the starting node and the set of new
neighbours must exclude the nodes already considered in the neighbours of previous
nodes. The obtained tree is proven to have exactly one leaf for each size $k$ subgraph.
The most important facts about ESU are that the ESU-tree can be efficiently sampled
to extract subgraphs without any bias and that its paths can be randomly explored to
estimate the total number of size $k$ subgraphs. Thanks to the ESU-tree, full enumeration of
the subgraphs can be avoided to estimate frequency and an efficient unbiased
probabilistic sampling, called RAND-ESU, can be used to randomly skip subgraphs. The
idea behind RAND-ESU is to call the recursive function with a given probability. It corresponds to exploring only a section of the ESU-tree, but in a way such that each leaf can be
reached with the same probability. RAND-ESU uses the $F1$ frequency.

\section{Data}
\label{data}

Data are obtained from the UN Comtrade website\footnote{http://comtrade.un.org}, querying the online system. The
metadata used for motif detection are: commodities (all), reporter code (all), partner code
(all), year (corresponding to the performed tests), trade flow (import/export). The total trade
value in dollars is the weight measuring the strength of connections. Subsequently, the
data are filtered by isolating only directional information in the graph (reporter code and
partner code) representing the flow of commodities between the different partners. When
more than one year is considered, the weights representing flow values in each year have
been added to obtain a cumulative value:

\begin{equation}
S_{ij}=\sum_{years}C_{ij}\, , 
\end{equation}

where $S_{ij}$ is the cumulative flow of commodities in US dollars from Country $C_{i}$ to Country $C_{j}$ over the considered years. One graph is obtained representing all the considered years
and $S_{ij}$ is the weight of the edges between nodes in this graph.
Theoretically if the Country $C_{i}$ has an export connection with the Country $C_{j}$ then the
latter should have an import connection with the former with the same weight and opposite
direction. Due to the differences in reporting procedures and data sources, in practice this
equality does not hold: the import graph cannot be obtained simply switching the direction
of edges of the export graph and vice versa. While this lack of homogeneity is a generally
acknowledged fact, the common practice of averaging the two values does not work for
macroscopic differences. In the performed experiments, we have found graphs of import
and export of the same year with different number of nodes and edges, together with
marked differences in weights very unlikely to be averaged out. Furthermore, it must be
observed that in the search for motifs, switching the direction of edges in the network
changes also the overrepresented motifs, and there is not a straightforward
correspondence between a motif and its reverse. For these reasons, only the directed
graph corresponding to export relationships is considered from now on.

\section{Motif Analysis of WTW}
\label{MAWTW}

Topological properties depend by definition upon the structure of the network, and
edge weights do not influence them, so that weak connections have the same relevance of
strong connections. To highlight how the strength of connections can influence the
topology, in terms of correlations or structural differences among strongly and weakly
connected nodes, a network split is proposed in this paper.
The first step is to split each graph in two subgraphs, called ``inliers'' and ``outliers'',
based on the edge weights. Table \ref{data} reports threshold values adopted to cut the WTW
graphs. Given the highly skewed distribution of weights, there is an evident discontinuity
threshold in the histograms, that after a number of trials have been found optimal with
10\, 000 bins. 



Analyzing separately the inliers and outliers graphs with RAND-ESU produces Table \ref{r1}, where the average abundance and the p-value of significant motifs of size 3 on the
export WTW from year 2004 to year 2006 are reported.

\begin{table}[h]
\centering
\caption{Results on the three-year period 2004-2006. Start year, end year, type, motif ID, percentage of
abundance and P-value are reported.}
\label{r1}
\scriptsize
\begin{tabular}{cccccc}
\hline
\textbf{Syear} & \textbf{Eyear} &  \textbf{Type}  & \textbf{IDM}  & \textbf{Percentage}  & \textbf{P}\\
\hline
2004 & 2006 & Inliers & 6 & 14,15\% & 0,990\\
2004 & 2006 & Inliers & 12 & 4,31\% & 0,000\\
2004 & 2006 & Inliers & 14 & 21,66\% & 0,003\\
2004 & 2006 & Inliers & 36 & 4,40\% & 0,002\\
2004 & 2006 & Inliers & 46 & 9,44\% & 0,020\\
2004 & 2006 & Inliers & 164 & 2,30\% & 0,000\\
2004 & 2006 & Inliers & 238 & 4,64\% & 0,004\\
2004 & 2006 & Outliers & 6 & 9,70\% & 0,000\\
2004 & 2006 & Outliers & 12 & 4,10\% & 0,000\\
2004 & 2006 & Outliers & 14 & 28,62\% & 0,008\\
2004 & 2006 & Outliers & 36 & 10,14\% & 0,999\\
2004 & 2006 & Outliers & 46 & 4,13\% & 0,000\\
2004 & 2006 & Outliers & 164 & 6,02\% & 0,005\\
2004 & 2006 & Outliers & 238 & 7,00\% & 0,000\\

\hline   
\end{tabular}
\end{table}

IDM is the code for the motif. The purpose was to highlight the
most evident features, so two filters were used: one on the significance (p-value less than
0.05) and another one on frequency (more than 0.05). Results for all motifs that met at
least one of these criteria are reported. As test case, data before, after and during the
financial crisis of 2007 are shown in tables.

\begin{figure} [!ht]
\centering
\includegraphics[scale=0.25]{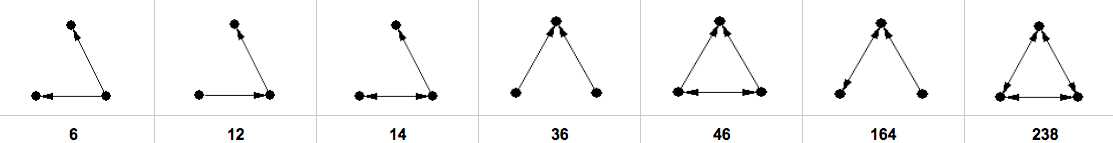}
\caption{Significant motifs distinct by id.}
\label{ms}
\end{figure}

\begin{table}[!ht]
\centering
\caption{Results on the year 2007.}
\label{r2}
\scriptsize
\begin{tabular}{cccccc}
\hline
\textbf{Syear} & \textbf{Eyear} &  \textbf{Type}  & \textbf{IDM}  & \textbf{Percentage}  & \textbf{P}\\
\hline

2007 & 2007 & Inliers & 6 & 15,91\% & 0,995\\
2007 & 2007 & Inliers & 12 & 5,60\% & 0,000\\
2007 & 2007 & Inliers & 14 & 22,37\% & 0,000\\
2007 & 2007 & Inliers & 36 & 4,80\% & 0,000\\
2007 & 2007 & Inliers & 46 & 8,14\% & 0,071\\
2007 & 2007 & Inliers & 164 & 2,50\% & 0,000\\
2007 & 2007 & Inliers & 238 & 2,10\% & 0,003\\
2007 & 2007 & Outliers & 6 & 10,95\% & 0,000\\
2007 & 2007 & Outliers & 12 & 4,37\% & 0,000\\
2007 & 2007 & Outliers & 14 & 29,29\% & 0,006\\
2007 & 2007 & Outliers & 36 & 9,20\% & 1,000\\
2007 & 2007 & Outliers & 46 & 4,50\% & 0,000\\
2007 & 2007 & Outliers & 164 & 5,43\% & 0,001\\
2007 & 2007 & Outliers & 238 & 6,88\% & 0,000\\

\hline   
\end{tabular}
\end{table}

\begin{table}[!ht]
\centering
\caption{Results on the three-year period 2008-2010.}
\label{r3}
\scriptsize
\begin{tabular}{cccccc}
\hline
\textbf{Syear} & \textbf{Eyear} &  \textbf{Type}  & \textbf{IDM}  & \textbf{Percentage}  & \textbf{P}\\
\hline

2008 & 2010 & Inliers & 6 & 13,64\% & 0,979\\
2008 & 2010 & Inliers & 12 & 4,39\% & 0,000\\
2008 & 2010 & Inliers & 14 & 21,30\% & 0,000\\
2008 & 2010 & Inliers & 36 & 4,29\% & 0,000\\
2008 & 2010 & Inliers & 46 & 9,94\% & 0,036\\
2008 & 2010 & Inliers & 164 & 21,14\% & 0,000\\
2008 & 2010 & Inliers & 238 & 5,51\% & 0,003\\
2008 & 2010 & Outliers & 6 & 10,61\% & 0,000\\
2008 & 2010 & Outliers & 12 & 4,30\% & 0,000\\
2008 & 2010 & Outliers & 14 & 29,69\% & 0,007\\
2008 & 2010 & Outliers & 36 & 9,15\% & 1,000\\
2008 & 2010 & Outliers & 46 & 4,52\% & 0,000\\
2008 & 2010 & Outliers & 164 & 5,04\% & 0,005\\
2008 & 2010 & Outliers & 238 & 6,94\% & 0,000\\

\hline   
\end{tabular}
\end{table}

\begin{table}[!ht]
\centering
\caption{Results on the three-year period 2008-2010.}
\label{r4}
\scriptsize
\begin{tabular}{ccccc}
\hline
\textbf{IDM} &  \textbf{Type}  & \textbf{Before}  & \textbf{During}  & \textbf{After}\\
\hline

6 & Inliers & -- & -- & --\\
6 & Outliers & 9,70\% & 10,95\% & 10,61\%\\
12 & Inliers & 4,31\% & 5,60\% & 4,39\%\\
12 & Outliers & 4,10\% & 4,37\% & 4,30\%\\\hline
 & \textbf{Difference:} & \textbf{0,21\%} & \textbf{1,23\%} & \textbf{0,09\%}\\\hline
14 & Inliers & 21,66\% & 22,37\% & 21,30\%\\
14 & Outliers & 28,62\% & 29,29\% & 29,69\%\\\hline
 & \textbf{Difference:} & \textbf{-6,96\%} & \textbf{-6,92\%} & \textbf{-8,39\%}\\\hline
36 & Inliers & 4,40\% & 4,80\% & 4,29\%\\
36 & Outliers & -- & -- & --\\
46 & Inliers & 9,44\% & -- & 9,94\%\\
46 & Outliers & 4,13\% & 4,50\% & 4,52\%\\\hline
 & \textbf{Difference:}& \textbf{5,31\%} & \textbf{--} & \textbf{5,42\%}\\\hline
164 & Inliers & 2,30\% & 2,50\% & 21,14\%\\
164 & Outliers & 6,02\% & 5,43\% & 5,04\%\\\hline
 & \textbf{Difference:} & \textbf{-3,72\%} & \textbf{-2,93\%} & \textbf{16,10\%}\\\hline
238 & Inliers & 4,64\% & 2,10\% & 5,51\%\\
238 & Outliers & 7,00\% & 6,88\% & 6,94\%\\\hline
 & \textbf{Difference:} & \textbf{-2,36\%} & \textbf{-4,78\%} & \textbf{-1,43\%}\\

\hline   
\end{tabular}
\end{table}

Table `Results on the three-year period 2008-2010' gives an overview of the differences before, during and after the 2007 financial
crisis. The most evident difference, ca 16$\%$, is between outliers and inliers in motif 164,
that after the crisis has a marked increase in abundance. The second major difference, ca
8$\%$, is in motif 14, that is consistently more abundant in outliers than in inliers in all tested
cases. Motif 46 before and after the crisis and motif 238 (fully connected triad) during the
crisis are the ones that show the third major difference between inliers and outliers, in the
order of 5$\%$. During the year 2007 the motif 46 (uplinked mutual dyad) becomes non
significant in the inliers and the difference between outliers and inliers in pattern 238
becomes more evident.
As the standard deviations for all computed frequencies are very narrow (in the order of
1/10\, 000), all differences are statistically significant.

\section{Conclusions}
Communication models should consider overrepresented motifs to fully exploit the communication potential of a network for a given topology, identifying channels or infrastructures to be strengthened and choosing the most suitable message routing schema or network protocol. 
Based on the test case of the World Trade Web in the period 2003-2010,
differences in abundance and in the number of significant motifs have been highlighted
that help to characterize the WTW and to understand the evolution patterns of trade
relationship among Countries in terms of volume and time. This represents an hemblematic model for a networks of interactions of human organizations.
It has been shown that overrepresented motifs change with the strength of the
connections and hence splitting a network is a viable way to gain insight into its structure
and dynamics. 
Future work is in studying motifs of size greater than 3 and the induced effect of self-organization.







%


\end{document}